\documentclass[11pt,a4paper]{article}
\usepackage[T1]{fontenc}
\usepackage[utf8]{inputenc}
\usepackage{lmodern}
\usepackage[margin=23mm]{geometry}
\usepackage{amsmath,amssymb,amsthm}
\usepackage{microtype}
\usepackage{xcolor}
\usepackage{tikz}
\usepackage{caption}
\usepackage{placeins}
\usepackage{cite}
\definecolor{ink}{HTML}{183247}
\definecolor{matchblue}{HTML}{165F9F}
\definecolor{coverteal}{HTML}{087B70}
\definecolor{softgray}{HTML}{788592}
\definecolor{setalpha}{HTML}{79489B}
\definecolor{setbeta}{HTML}{A65E10}
\definecolor{setchi}{HTML}{087E8B}
\colorlet{Panel}{matchblue!5}
\tikzset{
  vertex/.style={circle,draw=ink,fill=white,minimum size=7.2mm,
    inner sep=0pt,font=\fontsize{8.5}{10}\selectfont},
  sat/.style={vertex,fill=coverteal!18,draw=coverteal,line width=.7pt},
  graph edge/.style={draw=softgray!75,line width=.55pt},
  matching/.style={draw=matchblue,line width=2.2pt},
  connecting edge/.style={draw=softgray!35,line width=.35pt},
  vtx/.style={vertex},
  edge/.style={graph edge},
  conn/.style={connecting edge},
  match/.style={matching}
}

\tikzset{
  vertex/.style={circle,draw=ink,fill=white,minimum size=7.2mm,inner sep=0pt,font=\fontsize{8.5}{10}\selectfont},
  graph edge/.style={draw=softgray!75,line width=.55pt},
  join edge/.style={draw=softgray!35,line width=.35pt},
  note/.style={font=\small,align=center,text=ink}
}

\usepackage[colorlinks=true,linkcolor=matchblue,citecolor=matchblue,
  urlcolor=matchblue]{hyperref}
\usepackage{cleveref}
\theoremstyle{plain}
\newtheorem{theorem}{Theorem}

\newtheorem{corollary}{Corollary}
\newtheorem{claim}{Claim}
\newenvironment{claimproof}[1][Proof of Claim~\theclaim]
  {\begin{proof}[#1]}
  {\end{proof}}
\newcommand{\EDS}{\textnormal{\textsc{Edge Dominating Set}}}
\newcommand{\ThreePartition}{\textnormal{\textsc{3-Partition}}}
\newcommand{\XThreeC}{\textnormal{\textsc{Exact Cover by 3-Sets}}}
\hypersetup{
  pdftitle={Edge Dominating Set is NP-complete on Chordal Graphs},
  pdfauthor={Tesshu Hanaka},
  pdfsubject={NP-completeness on P8-free chordal graphs and P12-free strongly chordal graphs}
}
\title{\bfseries Edge Dominating Set is NP-complete\\ on Chordal Graphs\thanks{This work is partially supported by JSPS KAKENHI Grant Numbers 
JP22H00513, 
JP25K03077, 
and JST, CRONOS, Japan Grant Number JPMJCS24K2.}}
\author{%
  Tesshu Hanaka\\[0.3em]
  {\small Kyushu University, Fukuoka, Japan}%
}
\date{}
\begin{document}
\maketitle
\begin{abstract}
In this paper, we show that \EDS{} is NP-complete on chordal graphs.
Moreover, the problem remains NP-complete even on strongly chordal graphs, a subclass of chordal graphs.
This resolves the question posed by Horton and Kilakos [{SIAM} J. Discret. Math.,
6(3):375--387, 1993] and highlighted by Manlove [Discret. Appl. Math., 91(1--3):155--175, 1999].
\end{abstract}

\section{Introduction}\label{sec:intro}

In a graph $G=(V,E)$, an edge \emph{dominates} itself and every edge sharing an endpoint with it.
An \emph{edge dominating set} is a set of edges that dominates every edge of the graph.
Given a graph $G$ and a nonnegative integer $k$, \EDS{} asks whether $G$ has an edge dominating set of size at most $k$.

A \emph{matching} is a set of edges no two of which share an endpoint.
A matching is \emph{maximal} if no edge can be added to it while preserving the matching property.
Given a graph $G$ and a nonnegative integer $k$, \textsc{Minimum Maximal Matching} asks whether $G$ has a maximal matching of size at most $k$.
A matching is maximal exactly when it is an edge dominating set, and every graph has a minimum edge dominating set that is a matching~\cite{YG80}.
Thus, the minimum size of an edge dominating set equals the minimum size of a maximal matching.
Yannakakis and Gavril~\cite{YG80} prove that \EDS{} and the equivalent problem \textsc{Minimum Maximal Matching} are NP-complete.

Several studies have examined the complexity of \EDS{} on special graph classes.
Yannakakis and Gavril~\cite{YG80} show that \EDS{} is NP-complete on planar graphs of maximum degree at most three and on bipartite graphs of maximum degree at most three.
Horton and Kilakos~\cite{HK93} also prove that \EDS{} is NP-complete on planar bipartite graphs and their line and total graphs, as well as on perfect claw-free graphs and planar cubic graphs.
Demange and Ekim~\cite{DE08} prove that \textsc{Minimum Maximal Matching}, and hence \EDS{}, is NP-complete on $k$-regular bipartite graphs for every fixed $k\ge3$.
\EDS{} admits linear-time algorithms on trees~\cite{MH77}, series-parallel graphs~\cite{RP88}, and block graphs~\cite{HC95}.
A graph is \emph{chordal} if it contains no induced cycle of length at least four, and it is \emph{cotriangulated} if its complement is chordal.
Srinivasan et~al.~\cite{S95} give polynomial-time algorithms for \EDS{} on cotriangulated graphs and bipartite permutation graphs.
Horton and Kilakos~\cite{HK93} show that \EDS{} is polynomial-time solvable on claw-free chordal graphs, a superclass of proper interval graphs.
They noted that its complexity on chordal graphs was unknown, and Manlove~\cite{Manlove} later highlighted this question.

\subsection{Our contribution}

In this paper, we prove that \EDS{} is NP-complete on chordal graphs by a reduction from \XThreeC{} (Theorem~\ref{thm:chordal}).
This resolves the question left open by Horton and Kilakos~\cite{HK93} and later highlighted by Manlove~\cite{Manlove}.
By a reduction from \ThreePartition{}, we also prove that \EDS{} remains NP-complete on \emph{strongly chordal graphs}, a subclass of chordal graphs (Theorem~\ref{thm:strongchordal}).

Both reductions also bound the number of vertices in any induced path in the constructed graphs.
Let $P_t$ denote the path on $t$ vertices. A graph is \emph{$P_t$-free} if it contains no induced $P_t$.
The first reduction produces $P_8$-free chordal graphs, and the second produces $P_{12}$-free strongly chordal graphs.
Since every $P_t$ with $t\ge8$ contains an induced $P_8$, \EDS{} is NP-complete on $P_t$-free chordal graphs for each fixed $t\ge8$.
This partially answers a question of Milani\v{c}~\cite[Problem~2.1]{Milanic22} about the complexity of \EDS{} on $P_t$-free graphs for each fixed $t\ge5$.
Note that $P_4$-free graphs are exactly the cographs~\cite{CorneilLS81}, which have clique-width at most two~\cite{CourcelleO00}.
Hence, \EDS{} can be solved in polynomial time on $P_4$-free graphs~\cite{EGW01,KR03}.
To the best of our knowledge, the complexity of \EDS{} on $P_t$-free chordal graphs remains open for $t\in\{5,6,7\}$.

\subsection{Related work}

One problem related to \EDS{} is \textsc{Dominating Set}.
A \emph{dominating set} of a graph is a vertex set $D$ such that every vertex outside $D$ has a neighbor in $D$.
An edge dominating set of $G$ corresponds to a dominating set of the line graph of $G$.
\textsc{Dominating Set} is the problem of deciding whether a graph has a dominating set of size at most $k$ for a given integer $k$.
\textsc{Dominating Set} is NP-complete even on split graphs, and hence on chordal graphs~\cite{Bertossi}.
On strongly chordal graphs, \textsc{Dominating Set} is polynomial-time solvable~\cite{Farber84}.
Our result establishes NP-completeness for \EDS{} on the same class.

Other related problems are \textsc{Perfect Edge Dominating Set} and \textsc{Efficient Edge Dominating Set}.
Given a graph and an integer $k$, \textsc{Perfect Edge Dominating Set} asks whether the graph has an edge dominating set $F$ with $|F|\le k$ such that each edge not in $F$ shares an endpoint with exactly one edge of $F$.
\textsc{Efficient Edge Dominating Set} imposes the additional condition that no two edges in $F$ share an endpoint.
Both problems are NP-complete in general and are solvable in polynomial time on chordal graphs~\cite{LKT}.

A matching $M$ is \emph{2-maximal} if it is maximal and no single edge in $M$ can be replaced by two edges outside $M$ while preserving the matching property~\cite{Dean}.
Given a graph $G$ and an integer $k$, \textsc{Minimum 2-Maximal Matching} asks whether $G$ has a 2-maximal matching of size at most $k$.
Dean et~al.~\cite{Dean} prove that \textsc{Minimum 2-Maximal Matching} is NP-complete on chordal graphs.
As in our reduction in Section~\ref{sec:chordal}, their reduction uses a clique of element vertices with a leaf attached to each vertex.
Their reduction represents each triple by a triangle whose vertices are adjacent to all three elements of the triple.
In our chordal reduction, a path on seven vertices represents each triple.

\section{Preliminaries}\label{sec:preliminaries}
We use standard graph-theoretic notation.
Let $G=(V,E)$ be a simple undirected graph.
For a vertex $v$, we denote by $N_G(v)$ the neighborhood of $v$ and define $N_G[v]=N_G(v)\cup\{v\}$.
We write $[r]=\{1,\ldots,r\}$ for a positive integer $r$.
For $S\subseteq V(G)$, we denote by $G[S]$ the subgraph of $G$ induced by $S$, and we write $G-S=G[V(G)\setminus S]$.

For an ordering $v_1,\ldots,v_n$ of $V(G)$, let $G_i=G[\{v_i,\ldots,v_n\}]$ for each $i\in[n]$.
The ordering is a \emph{perfect elimination ordering} (PEO) if $N_{G_i}(v_i)$ is a clique for every $i\in[n]$.
A graph with a PEO is chordal~\cite{FG}.

The ordering is a \emph{strong elimination ordering} (SEO)~\cite{Farber83,BCD98,CFT93,SawadaS03} if, for every $i\in[n]$,
\begin{equation}\label{eq:seo-neighborhood-order}
 N_{G_i}[v_j]\subseteq N_{G_i}[v_k]
 \qquad\text{whenever }v_j,v_k\in N_{G_i}[v_i]\text{ and }j<k.
\end{equation}
Every SEO is a PEO.
Indeed, for any $v_j,v_k\in N_{G_i}(v_i)$ with $j<k$, we have $v_j\in N_{G_i}[v_j]\subseteq N_{G_i}[v_k]$, and hence $v_j$ and $v_k$ are adjacent.
A graph is \emph{strongly chordal} if and only if it admits an SEO~\cite{Farber83}.

Throughout this paper, we only show the NP-hardness of \EDS{} because the problem is in NP~\cite{YG80}.

\section{NP-completeness on chordal graphs}
\label{sec:chordal}
In this section, we show that \EDS{} is NP-complete on chordal graphs.
We also show that the constructed graphs are $P_8$-free.

\begin{theorem}\label{thm:chordal}
\EDS{} is NP-complete on chordal graphs.
\end{theorem}

\begin{proof}
We give a reduction from \XThreeC{}, which is NP-complete~\cite[Problem~SP2]{GJ79}.
An instance of \XThreeC{} consists of a set $X=\{x_1,\ldots,x_{3n}\}$ and a family $\mathcal S=\{S_1,\ldots,S_m\}$ of three-element subsets of $X$.
The question is whether some $n$ members of $\mathcal S$ partition $X$.

We construct a graph $G$ as follows.
We identify $X=\{x_1,\ldots,x_{3n}\}$ as the set of \emph{element vertices} in $G$.
Then we make $X$ a clique and attach a leaf $y_i$ to each $x_i$.
We call each edge $\{x_i,y_i\}$ a \emph{leaf edge}.
For each $S_j\in\mathcal S$, we add a path $P_j=z_{j,1}\cdots z_{j,7}$ and join each of its seven vertices to all three vertices in $S_j$.
These are the only edges between $P_j$ and the rest of the graph.
We refer to the edges within a path as \emph{path edges} and to the edges between a path and $X$ as \emph{connecting edges}.
The graph $G$ has $6n+7m$ vertices and $\binom{3n}{2}+3n+27m$ edges, and hence it can be constructed in polynomial time.

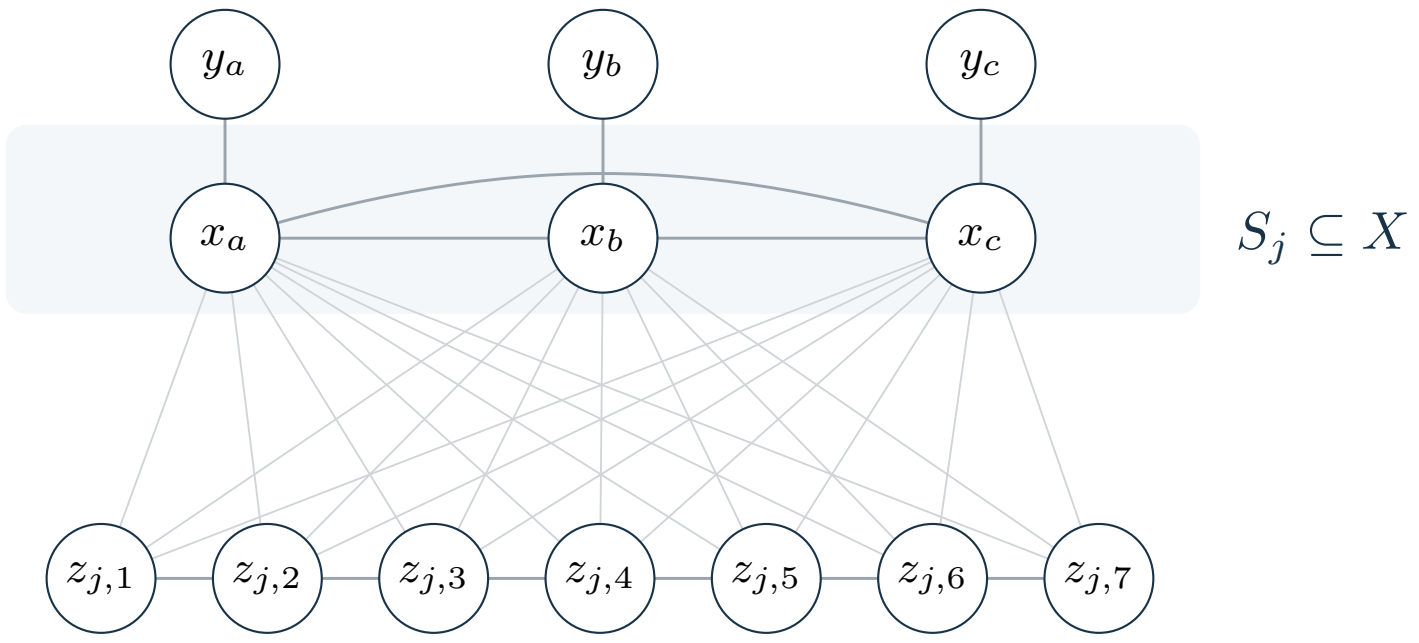
\begin{figure}
  \centering
  \begin{tikzpicture}[x=1cm,y=1cm]
  \fill[matchblue!5,rounded corners] (-.45,2.15) rectangle (7.45,3.4);
  \node[note,anchor=west] at (7.55,2.65) {$S_j\subseteq X$};
  \foreach \k/\xx in {a/1,b/3.5,c/6}{
    \coordinate (x\k) at (\xx,2.65);
    \coordinate (y\k) at (\xx,3.8);
  }
  \foreach \r in {0,...,6}{\coordinate (z\r) at ({.18+1.1*\r},.4);}
  \foreach \k in {a,b,c}{\foreach \r in {0,...,6}{\draw[join edge] (x\k)--(z\r);}}
  \draw[graph edge] (xa)--(xb)--(xc);
  \draw[graph edge] (xa) to[bend left=17] (xc);
  \foreach \k in {a,b,c}{\draw[graph edge] (x\k)--(y\k);}
  \foreach \r in {0,...,5}{\pgfmathtruncatemacro{\s}{\r+1}\draw[graph edge] (z\r)--(z\s);}
  \foreach \k in {a,b,c}{\node[vertex] at (x\k) {$x_{\k}$};\node[vertex] at (y\k) {$y_{\k}$};}
  \foreach \r in {0,...,6}{\pgfmathtruncatemacro{\lab}{\r+1}\node[vertex] at (z\r) {$z_{j,\lab}$};}
\end{tikzpicture}
\caption{The construction for $S_j=\{x_a,x_b,x_c\}$. Each of the seven path vertices is adjacent to all three elements of $S_j$; the faint lines represent these 21 connecting edges.
The whole set $X$ forms a clique, and each $x_i$ has exactly one leaf $y_i$ in the entire graph.}
\label{fig:construction}
\end{figure}

We first show that $G$ is chordal.
Consider the ordering that lists the leaves first, then the vertices of each path from $z_{j,1}$ to $z_{j,7}$, and finally the vertices of $X$.
This ordering is a PEO.
Indeed, when a path vertex is deleted, its remaining neighbors are the clique $S_j$ and, if present, the next path vertex, which is adjacent to all vertices of $S_j$.

We next show that $\mathcal S$ contains an exact cover if and only if $G$ has an edge dominating set of size at most $2m+n$.
Suppose that $\mathcal S$ contains an exact cover.
For each $S_j$ in the cover, we select three connecting edges that match the three elements of $S_j$ to $z_{j,2}$, $z_{j,4}$, and $z_{j,6}$.
For every other $S_j$, we select the two path edges $\{z_{j,2},z_{j,3}\}$ and $\{z_{j,5},z_{j,6}\}$.
Figure~\ref{fig:states} illustrates the two choices.
In both cases, the selected edges dominate all six edges of $P_j$.
Since the cover partitions $X$, every vertex of $X$ is an endpoint of a selected edge.
Thus, all edges incident to $X$ are also dominated.
The number of selected edges is $3n+2(m-n)=2m+n$.

\begin{figure}[tbp]
\centering
\begin{tikzpicture}[x=.96cm,y=.9cm]
\foreach \shift/\mode in {0/0,8/1}{
 \begin{scope}[xshift=\shift cm]
 \foreach \i in {1,...,7}{\coordinate (z\i) at (\i,0);}
 \foreach \i in {1,...,6}{\pgfmathtruncatemacro{\j}{\i+1}\draw[edge] (z\i)--(z\j);}
 \ifnum\mode=0
  \draw[match] (z2)--(z3);\draw[match] (z5)--(z6);
  \foreach \i in {1,4,7}{\node[vtx] at (z\i) {\i};}
  \foreach \i in {2,3,5,6}{\node[sat] at (z\i) {\i};}
  \node[font=\small] at (4,1.75) {Two path edges};
 \else
  \foreach \i/\lab in {2/p,4/q,6/s}{
   \coordinate (x\i) at (\i,1.13);\draw[match] (z\i)--(x\i);
   \node[sat] at (x\i) {$x_{\lab}$};\node[sat] at (z\i) {\i};}
  \foreach \i in {1,3,5,7}{\node[vtx] at (z\i) {\i};}
  \node[font=\small] at (4,1.75) {Three connecting edges};
 \fi
  \node[font=\footnotesize] at (4,-.65) {No path edge connects two white vertices};
 \end{scope}
}
\end{tikzpicture}
\caption{The two choices for the path gadgets. A triple outside the chosen partition uses two path edges (left), and a triple in the partition uses three connecting edges (right). Both choices dominate all six path edges. Numbers indicate positions on the path. Blue edges are selected, and green vertices are their endpoints. The other connecting edges and the edges inside $X$ are omitted.}
\label{fig:states}
\end{figure}
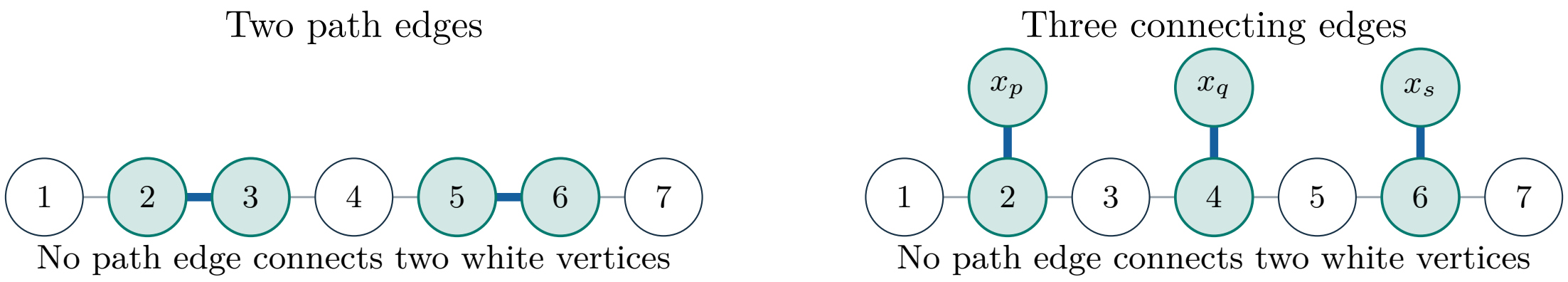

Conversely, suppose that $G$ has an edge dominating set $F$ with $|F|\le2m+n$.
Let $a$ be the number of paths $P_j$ such that $F$ contains a connecting edge incident to $P_j$.
We also denote by $t$ the number of edges of $F$ that lie inside $X$ or are leaf edges.
For each path, only its path edges and the connecting edges incident to it can dominate its six path edges.
A path edge dominates at most three path edges, and a connecting edge dominates at most two.
Hence, at least two edges of $F$ are incident to the vertices of each path.
Moreover, if one of them is a connecting edge, then at least three such edges are needed.
Summing over all paths, we obtain $|F|\ge 2(m-a)+3a+t=2m+a+t$.
Since $|F|\le2m+n$, we have $a+t\le n$.

Since every leaf edge $\{x_i,y_i\}$ must be dominated, every vertex $x_i\in X$ is incident to an edge of $F$.
For each of the $a$ paths $P_j$ counted above, every connecting edge in $F$ incident to $P_j$ has its endpoint in $X$ among the three vertices of $S_j$.
Thus, the connecting edges in $F$ are incident to at most $3a$ distinct vertices of $X$.
Each of the $t$ edges in $F$ that lie inside $X$ or are incident to a leaf has at most two endpoints in $X$.
Since every vertex of $X$ is incident to an edge of $F$ and path edges have no endpoints in $X$, we obtain
\[
  3n\le 3a+2t=3(a+t)-t\le 3n-t.
\]
It follows that $t=0$ and $a=n$.
Since $t=0$, every vertex of $X$ is incident to a connecting edge of $F$.
Thus, the triples corresponding to the $n$ paths counted by $a$ cover all $3n$ elements.
Since each triple has size three, these triples are pairwise disjoint and form an exact cover.
\end{proof}

\begin{corollary}\label{cor:p8-free}
\EDS{} is NP-complete on $P_8$-free chordal graphs.
\end{corollary}
\begin{proof}
Let $G$ be a graph constructed in the proof of Theorem~\ref{thm:chordal}, and let $Q$ be an induced path in $G$.
If $Q$ avoids $X$, it lies in a component of $G-X$ and has at most seven vertices.
Otherwise, $Q$ contains at least one vertex in $X$.
Since $X$ is a clique, $Q$ cannot contain more than two vertices of $X$. Moreover, any two vertices of $Q$ in $X$ must be consecutive on $Q$. Otherwise, the edge between them would be a chord of $Q$.
Consequently, the vertices of $Q$ outside $X$ form at most two segments, each lying in a single component of $G-X$, either a leaf or a path $P_j$.
If a segment lies in $P_j$, the vertex of $X$ next to the segment on $Q$ belongs to $S_j$ and is adjacent to every vertex of $P_j$.
A second vertex in the segment would therefore form a chord with that vertex of $X$.
Hence, each segment contains at most one vertex, so $Q$ has at most two vertices outside $X$ and $|V(Q)|\le 4$.
Thus, $G$ is $P_8$-free.
Since $G$ is also chordal, the reduction in the proof of Theorem~\ref{thm:chordal} proves the corollary.
\end{proof}

\section{NP-completeness on strongly chordal graphs}
\label{sec:strong}
In this section, we show that \EDS{} remains NP-complete on strongly chordal graphs.
In the reduction of Section~\ref{sec:chordal}, all vertices of a path have the same three neighbors in the clique $X$.
We now assign numerical thresholds to the path vertices.
The resulting neighborhoods in $X$ are nested, which allows us to give a strong elimination ordering.
Finally, we show that the constructed graphs are $P_{12}$-free.
\begin{theorem}\label{thm:strongchordal}
\EDS{} is NP-complete on strongly chordal graphs.
\end{theorem}

\begin{proof}
We give a reduction from \ThreePartition{}, which is NP-complete~\cite{GJ78,GJ79}.
An instance of \ThreePartition{} consists of a sequence $(a_1,\ldots,a_{3n})$ of $3n$ positive integers and a positive integer $B$ such that $\sum_{i=1}^{3n}a_i=nB$ and $B/4<a_i<B/2$ for every $i\in[3n]$.
The question is whether $[3n]$ can be partitioned into $n$ triples, each with total value $B$.
Equal values at different indices are treated as distinct items.
Without loss of generality, we assume that $a_1\ge a_2\ge\cdots\ge a_{3n}$.
Let $\mathcal T=\{T_1,\ldots,T_m\}$ be the family of all three-element subsets $T$ of $[3n]$ with $\sum_{i\in T}a_i=B$.
For each $T_j$, let $\alpha_j\ge\beta_j\ge\gamma_j$ be the values associated with its three indices, in nonincreasing order.
Thus, $\alpha_j+\beta_j+\gamma_j=B$.

We construct a graph $G$ as follows.
First, we create a clique $X=\{x_1,\ldots,x_{3n}\}$ of \emph{element vertices} and attach a leaf $y_i$ to each $x_i$. We call each edge $\{x_i,y_i\}$ a \emph{leaf edge}.
Then, for each $T_j\in\mathcal T$, we create a new path $P_j=z_{j,1}\cdots z_{j,7}$.
We assign to each vertex $z_{j,k}$ a threshold $\tau_{j,k}$, which is $\alpha_j$ for $k\in\{1,2\}$, $\beta_j$ for $k\in\{3,4\}$, and $\gamma_j$ for $k\in\{5,6,7\}$.
By definition, $\tau_{j,k}\ge\tau_{j,k+1}$ for every $k\in[6]$.
For each $i\in[3n]$ and $k\in[7]$, we add the edge $\{x_i,z_{j,k}\}$ if and only if $a_i\ge\tau_{j,k}$.
The closed neighborhood of $x_i$ consists of $X$, its leaf $y_i$, and its neighbors on the paths $P_j$.
For any $i,h\in[3n]$ with $a_i\ge a_h$, every path vertex $z_{j,k}$ adjacent to $x_h$ satisfies $a_i\ge a_h\ge\tau_{j,k}$ and is therefore adjacent to $x_i$.
Hence,
\begin{equation}\label{eq:closed-neighborhood-inclusion}
 N_G[x_h]\setminus\{y_h\}\subseteq N_G[x_i]\setminus\{y_i\}.
\end{equation}

The graph $G$ has $6n+7m$ vertices and $O(n^2+nm)$ edges.
Since $m\le\binom{3n}{3}$, we can construct $G$ in polynomial time.
As in Section~\ref{sec:chordal}, we refer to the edges within a path as \emph{path edges} and to the edges between a path and $X$ as \emph{connecting edges}.
For each $j\in[m]$ and $k\in[7]$, let $A_{j,k}=N_G(z_{j,k})\cap X=\{x_i\in X:a_i\ge\tau_{j,k}\}$.
By the definition of the thresholds, $A_{j,1}=A_{j,2}\subseteq A_{j,3}=A_{j,4}\subseteq A_{j,5}=A_{j,6}=A_{j,7}$ for every $j\in[m]$.
Figure~\ref{fig:threshold-interior} illustrates the construction.

\begin{figure}[tbp]
\centering
\begin{tikzpicture}[x=1.6cm,y=1cm,
  conn/.style={draw=black!40,line width=.55pt},
  edge/.style={draw=ink,line width=.7pt}]
\fill[Panel,rounded corners=2mm](-.35,1.6) rectangle (6.35,2.5);
\foreach \k/\thr in {0/11,1/11,2/10,3/10,4/9,5/9,6/9}{
 \foreach \pos/\val in {0/12,1.2/11,2.4/10,3.6/10,4.8/9,6/8}{
  \ifnum\val<\thr\else\draw[conn](\pos,2.02)--(\k,0);\fi}
}
\draw[setchi,densely dotted,line width=.9pt,rounded corners=1.5mm](-.62,1.01) rectangle (5.42,2.64);
\draw[setbeta,dashed,line width=.8pt,rounded corners=1.5mm](-.50,1.13) rectangle (4.10,2.53);
\draw[setalpha,line width=.8pt,rounded corners=1.5mm](-.38,1.25) rectangle (1.58,2.42);
\foreach \k in {0,...,5}{\pgfmathtruncatemacro{\h}{\k+1}\draw[edge](\k,0)--(\h,0);}
\foreach \pos/\idx/\val in {0/1/12,1.2/2/11,2.4/3/10,3.6/4/10,4.8/5/9,6/6/8}{
 \draw[edge](\pos,2.02)--(\pos,3.05);
 \node[vtx] at (\pos,2.02){$x_{\idx}$};\node[vtx] at (\pos,3.05){$y_{\idx}$};
 \node[font=\footnotesize,fill=white,inner sep=1pt] at (\pos,1.47){$a_{\idx}=\val$};
}
\foreach \k/\thr in {0/11,1/11,2/10,3/10,4/9,5/9,6/9}{
 \pgfmathtruncatemacro{\h}{\k+1}
 \node[vtx] at (\k,0){$z_{j,\h}$};\node[font=\footnotesize] at (\k,-.55){$\tau=\thr$};
}
\node[font=\small,align=center] at (3,3.9){$B=30,\quad T_j=\{2,3,5\}$\\[2pt]$(\alpha_j,\beta_j,\gamma_j)=(11,10,9)$};
\draw[setalpha,line width=.8pt](-.3,-.78)--(-.3,-.94)--(1.3,-.94)--(1.3,-.78);
\draw[setbeta,line width=.8pt](1.7,-.78)--(1.7,-.94)--(3.3,-.94)--(3.3,-.78);
\draw[setchi,line width=.8pt](3.7,-.78)--(3.7,-.94)--(6.3,-.94)--(6.3,-.78);
\node[font=\footnotesize,text=setalpha,align=center,anchor=north] at (.5,-1.02){$A_{j,1}=A_{j,2}$\\[2pt]$=\{x_1,x_2\}$};
\node[font=\footnotesize,text=setbeta,align=center,anchor=north] at (2.5,-1.02){$A_{j,3}=A_{j,4}$\\[2pt]$=\{x_1,x_2,x_3,x_4\}$};
\node[font=\footnotesize,text=setchi,align=center,anchor=north] at (5,-1.02){$A_{j,5}=A_{j,6}=A_{j,7}$\\[2pt]$=\{x_1,x_2,x_3,x_4,x_5\}$};
\node[font=\small] at (3,-2.38){$A_{j,1}\subseteq A_{j,3}\subseteq A_{j,5}$};
\end{tikzpicture}
\caption{Part of the graph constructed from the \ThreePartition\ instance $(a_1,\ldots,a_6)=(12,11,10,10,9,8)$ with $n=2$ and $B=30$. The figure focuses on the path $P_j$ corresponding to $T_j=\{2,3,5\}$, whose values are $(\alpha_j,\beta_j,\gamma_j)=(11,10,9)$, and on the inclusions $A_{j,1}\subseteq A_{j,3}\subseteq A_{j,5}$ among the neighborhoods of its vertices in $X$. The purple, orange, and teal outlines enclose $A_{j,1}$, $A_{j,3}$, and $A_{j,5}$, respectively, and the brackets in the same colors mark the corresponding path positions. All element vertices and their leaves are shown, whereas the edges inside $X$, all other paths, and their incident edges are omitted.}
\label{fig:threshold-interior}
\end{figure}
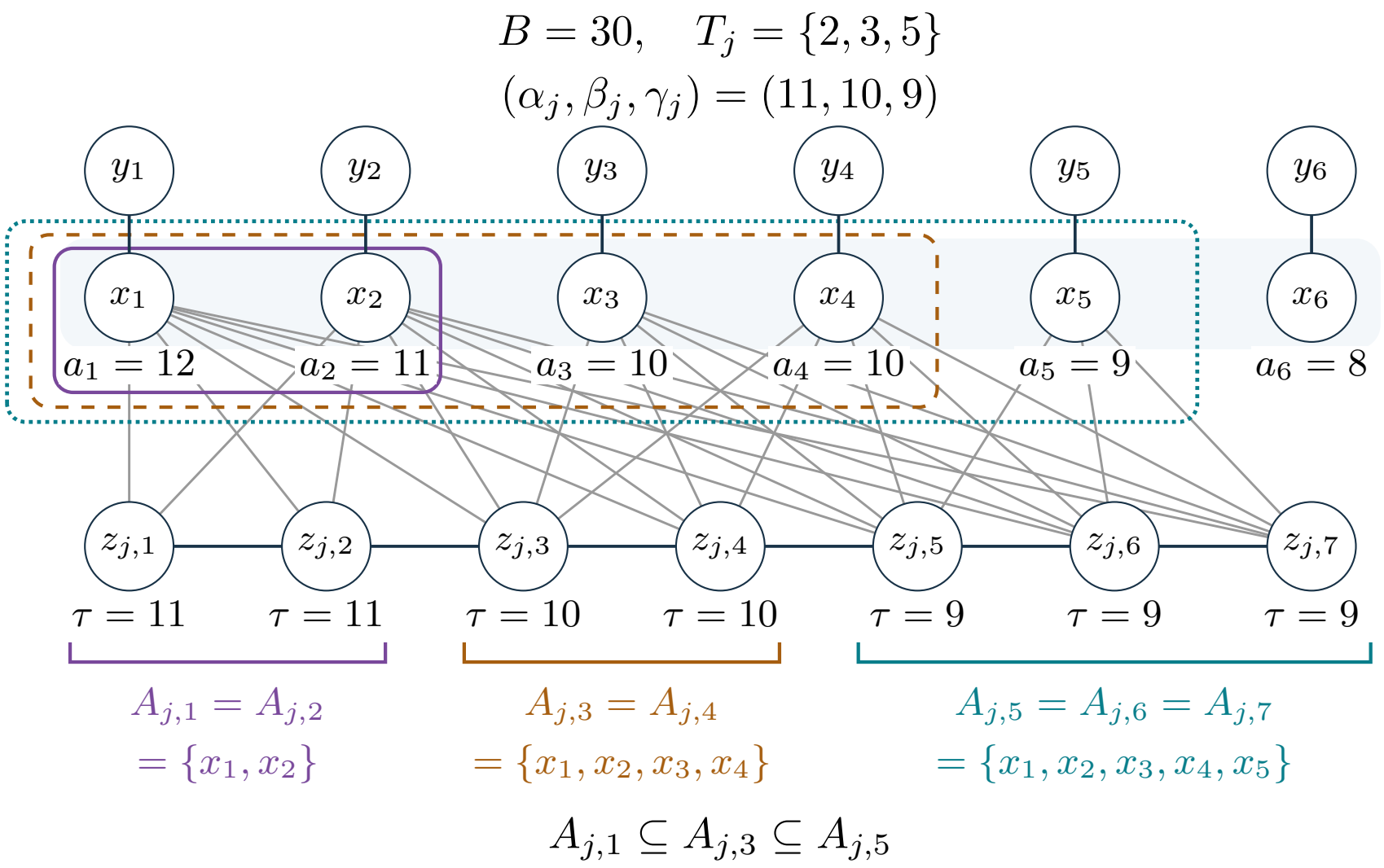

\begin{claim}
  The graph $G$ is strongly chordal.
\end{claim}
\begin{claimproof}
Let $v_1,\ldots,v_N$ be the ordering of $V(G)$ that lists $y_1,\ldots,y_{3n}$ first, then the paths $P_1,\ldots,P_m$, each in the order $z_{j,1},\ldots,z_{j,7}$, and finally $x_{3n},\ldots,x_1$.
For each $r\in[N]$, let $G_r=G[\{v_r,\ldots,v_N\}]$.
We show that this ordering satisfies~\eqref{eq:seo-neighborhood-order}.
That is, we show that $N_{G_r}[u]\subseteq N_{G_r}[v]$ whenever $u,v\in N_{G_r}[v_r]$ and $u$ precedes $v$ in the ordering.
We consider the three types of $v_r$ in the order of elimination: leaves, path vertices, and element vertices.

First, suppose that $v_r=y_i$.
Then $N_{G_r}[y_i]=\{y_i,x_i\}$, and $y_i$ precedes $x_i$.
Hence, the only pair to check is $(u,v)=(y_i,x_i)$, for which
$N_{G_r}[y_i]=\{y_i,x_i\}\subseteq N_{G_r}[x_i]$.

Next, suppose that $v_r=z_{j,k}$.
All leaves and the vertices $z_{j,1},\ldots,z_{j,k-1}$ have been eliminated, whereas all vertices of $X$ remain.
Thus,
\[
 N_{G_r}[z_{j,k}]=
 \begin{cases}
  \{z_{j,k},z_{j,k+1}\}\cup A_{j,k}, & k<7,\\
  \{z_{j,7}\}\cup A_{j,7}, & k=7.
 \end{cases}
\]
Let $u,v\in N_{G_r}[z_{j,k}]$ be such that $u$ precedes $v$.
Since every path vertex precedes every vertex of $X$, there are three cases.

If $u,v\in V(P_j)$, then necessarily $k<7$, $u=z_{j,k}$, and $v=z_{j,k+1}$.
Since $u$ and $v$ are adjacent and $A_{j,k}\subseteq A_{j,k+1}$, we have
$N_{G_r}[u]=\{u,v\}\cup A_{j,k}\subseteq N_{G_r}[v]$.

If $u\in V(P_j)$ and $v=x_i\in A_{j,k}$, then every vertex of $N_{G_r}[u]$ belongs to $X$ or to the remaining part of $P_j$.
For every $h\in\{k,\ldots,7\}$, we have $a_i\ge\tau_{j,k}\ge\tau_{j,h}$.
Thus, $x_i$ is adjacent to every remaining vertex of $P_j$.
Since $X$ is a clique, we have
\[
 N_{G_r}[u]\subseteq X\cup\{z_{j,k},\ldots,z_{j,7}\}
 \subseteq N_{G_r}[x_i]=N_{G_r}[v].
\]

If $u=x_i$ and $v=x_h$ both belong to $X$, then $i>h$ because the vertices of $X$ are listed in decreasing order of index.
Hence, $a_i\le a_h$.
Suppose that a remaining path vertex $z_{\ell,d}$ is adjacent to $x_i$.
By the definition of the connecting edges, we have $a_i\ge\tau_{\ell,d}$, and hence $a_h\ge a_i\ge\tau_{\ell,d}$.
Thus, $G$ also contains the edge $\{x_h,z_{\ell,d}\}$.
Consequently, every remaining path vertex adjacent to $x_i$ is also adjacent to $x_h$.
Since no leaves remain and both closed neighborhoods contain all of $X$, we have $N_{G_r}[u]\subseteq N_{G_r}[v]$.
Therefore, \eqref{eq:seo-neighborhood-order} holds whenever $v_r$ is a path vertex.

Finally, suppose that $v_r\in X$.
Since all leaves and path vertices have been eliminated, $G_r$ is a clique.
Thus, $N_{G_r}[u]=N_{G_r}[v]=V(G_r)$ for all remaining vertices $u$ and $v$, and~\eqref{eq:seo-neighborhood-order} holds.

Therefore, $v_1,\ldots,v_N$ is an SEO, and $G$ is strongly chordal.
\end{claimproof}

\begin{claim}\label{claim:eds-3partition}
The instance of \ThreePartition\ has a solution if and only if $G$ has an edge dominating set of size at most $2m+n$.
\end{claim}
\begin{claimproof}
First, suppose that the instance of \ThreePartition\ has a solution.
For each $T_j=\{p,q,s\}$ in the solution with $p<q<s$, we have $(a_p,a_q,a_s)=(\alpha_j,\beta_j,\gamma_j)$, and hence
\[
 a_p=\alpha_j=\tau_{j,2},\qquad
 a_q=\beta_j=\tau_{j,4},\qquad
 a_s=\gamma_j=\tau_{j,6}.
\]
Recall that $G$ contains the connecting edge $\{x_i,z_{j,k}\}$ if and only if $a_i\ge\tau_{j,k}$.
Thus, the connecting edges $\{x_p,z_{j,2}\}$, $\{x_q,z_{j,4}\}$, and $\{x_s,z_{j,6}\}$ exist, and we select these three edges.
For every other $T_j$, we select the path edges $\{z_{j,2},z_{j,3}\}$ and $\{z_{j,5},z_{j,6}\}$ (see Figure~\ref{fig:states}).
Let $F$ be the set of selected edges.
Then $|F|=3n+2(m-n)=2m+n$, and $F$ dominates all path edges.
Since the solution partitions $[3n]$, every element vertex is incident to a selected connecting edge.
Thus, all edges inside $X$, leaf edges, and connecting edges are also dominated.
Therefore, $G$ has an edge dominating set of size $2m+n$.

Conversely, suppose that $G$ has an edge dominating set $F$ with $|F|\le2m+n$.
We call the edges in $F$ \emph{selected edges}.
Since $y_i$ is a leaf and the edge $\{x_i,y_i\}$ must be dominated, every element vertex $x_i$ is incident to a selected edge.

For each path $P_j$, let $b_j$ and $c_j$ be the numbers of selected path edges of $P_j$ and selected connecting edges incident to $P_j$, respectively.
Only these edges can dominate the six path edges of $P_j$.
Since a path edge dominates at most three of them and a connecting edge dominates at most two, we have
\begin{equation}\label{eq:local}
 3b_j+2c_j\ge6.
\end{equation}
Let $c=\sum_{j=1}^m c_j$ be the total number of selected connecting edges, and let $t$ be the number of edges in $F$ that lie inside $X$ or are leaf edges.
Every selected edge is a path edge, a connecting edge, an edge inside $X$, or a leaf edge.
Thus, $|F|=\sum_{j=1}^m b_j+c+t$.
Dividing~\eqref{eq:local} by $3$ gives $b_j\ge2-\frac{2c_j}{3}$ for each $j$.
Consequently,
\[
\begin{aligned}
 |F|
 &=\sum_{j=1}^m b_j+c+t\\
 &\ge\sum_{j=1}^m\left(2-\frac{2c_j}{3}\right)+c+t\\
 &=2m-\frac{2c}{3}+c+t\\
 &=2m+\frac c3+t.
\end{aligned}
\]
Combining this lower bound with $|F|\le2m+n$, we obtain
\begin{equation}\label{eq:budget}
 c+3t\le3n.
\end{equation}

Each selected path edge has no endpoint in $X$, and each selected connecting edge has exactly one endpoint in $X$.
Each selected edge inside $X$ has two endpoints in $X$, and each selected leaf edge has exactly one endpoint in $X$.
Therefore, at most $c+2t$ distinct vertices of $X$ are endpoints of selected edges.
Since all $3n$ element vertices are incident to selected edges, we have $3n\le c+2t$.
By~\eqref{eq:budget}, we obtain
\[
 3n\le c+2t=(c+3t)-t\le3n-t.
\]
Since $t\ge0$, this implies $t=0$.
Then the inequalities $3n\le c+2t$ and $c+3t\le3n$ give $3n\le c\le3n$, that is, $c=3n$.

Substituting $t=0$ and $c=3n$ into the bounds on $|F|$, we obtain
$2m+n\le|F|\le2m+n$, and hence $|F|=2m+n$.
Therefore,
\[
 \sum_{j=1}^m b_j=|F|-c-t=(2m+n)-3n=2m-2n.
\]
It follows that
\[
 \sum_{j=1}^m(3b_j+2c_j-6)
 =3(2m-2n)+2(3n)-6m=0.
\]
By~\eqref{eq:local}, every summand is nonnegative, and hence every summand is zero.
Thus, $3b_j+2c_j=6$ for every $j$.
The only nonnegative integer solutions are $(b_j,c_j)=(2,0)$ and $(0,3)$.
Since $\sum_{j=1}^m c_j=3n$, exactly $n$ paths have $(b_j,c_j)=(0,3)$.

On a path with $(b_j,c_j)=(0,3)$, the three selected connecting edges must dominate the six path edges without overlap.
In particular, no selected connecting edge is incident to an endpoint of the path, since such an edge dominates only one path edge.
Thus, by considering the path edges from left to right, we see that the selected connecting edges are incident to $z_{j,2}$, $z_{j,4}$, and $z_{j,6}$.

Each of the $3n$ selected connecting edges has exactly one endpoint in $X$. Since every vertex of $X$ is incident to a selected edge and $t=0$, all $3n$ vertices of $X$ occur among these endpoints. Hence each occurs exactly once.
Thus, for each of the $n$ paths with $(b_j,c_j)=(0,3)$, the indices of the three element vertices joined to it by selected connecting edges form a triple, and these $n$ triples partition $[3n]$.
For such a path $P_j$, let $u_j$, $v_j$, and $w_j$ be the values of the element vertices joined to $z_{j,2}$, $z_{j,4}$, and $z_{j,6}$, respectively.
By construction, the edge $\{x_i,z_{j,k}\}$ exists only if $a_i\ge\tau_{j,k}$.
Applying this to the three selected connecting edges, we obtain
\[
 u_j\ge\tau_{j,2}=\alpha_j,\qquad
 v_j\ge\tau_{j,4}=\beta_j,\qquad
 w_j\ge\tau_{j,6}=\gamma_j,
\]
and hence $u_j+v_j+w_j\ge B$.
Since the $n$ triples partition $[3n]$, their total value is $\sum_{i=1}^{3n}a_i=nB$.
Thus, every triple has value exactly $B$, and each of the three inequalities above holds with equality.
Consequently, the triples of indices used by the selected paths form a solution of the \ThreePartition\ instance, even when equal values occur at different indices.
\end{claimproof}
\end{proof}

\begin{corollary}\label{cor:p12-free}
\EDS{} is NP-complete on $P_{12}$-free strongly chordal graphs.
\end{corollary}
\begin{proof}
Let $G$ be the graph constructed in the proof of Theorem~\ref{thm:strongchordal}, and let $Q$ be an induced path in $G$.
We show that $Q$ has at most 11 vertices.
If $Q$ contains no vertex of $X$, then $Q$ lies in a component of $G-X$, which is a leaf or a path $P_j$.
Thus, $Q$ has at most seven vertices in this case.

Suppose that $Q$ contains a vertex of $X$.
Since $X$ is a clique and $Q$ is an induced path, $Q$ contains at most two vertices of $X$, and if it contains two, they are consecutive on $Q$.
As in the proof of Corollary~\ref{cor:p8-free}, the vertices of $Q$ outside $X$ form at most two segments, each of which lies in a component of $G-X$.

We first show that every segment has at most five vertices.
A segment in a leaf has only one vertex.
Let $R$ be a segment in a path $P_j$, and let $x_i$ be the vertex of $X$ next to $R$ on $Q$.
Since $x_i$ has a neighbor in $P_j$ and $\gamma_j$ is the smallest threshold on $P_j$, we have $a_i\ge\gamma_j$ by the construction of $G$.
Thus, $x_i$ is adjacent to $z_{j,5}$, $z_{j,6}$, and $z_{j,7}$.
On the other hand, $x_i$ is adjacent to only one vertex of $R$, since $Q$ is an induced path.
Hence, $R$ contains at most one of $z_{j,5}$, $z_{j,6}$, and $z_{j,7}$, and thus at most five vertices of $P_j$.

We now bound the number of vertices of $Q$.
If $Q$ contains exactly one vertex of $X$, then $Q$ consists of this vertex and at most two segments, and hence it has at most $1+2\cdot5=11$ vertices.
In the remaining case, $Q$ contains two vertices of $X$.
Let $x_i$ and $x_h$ be these vertices with $a_i\ge a_h$.
Suppose that the segment next to $x_h$ starts with a vertex $v\ne y_h$.
Then $v x_h x_i$ is a subpath of $Q$.
By~\eqref{eq:closed-neighborhood-inclusion}, $v$ is adjacent to $x_i$, so $\{x_i,v\}$ is a chord of $Q$.
Thus, the segment next to $x_h$, if it exists, consists of the leaf $y_h$.
Therefore, $Q$ has at most $5+2+1=8$ vertices in this case.

From the above, every induced path in $G$ has at most 11 vertices, and hence $G$ is $P_{12}$-free.
Since $G$ is also strongly chordal, the reduction in the proof of Theorem~\ref{thm:strongchordal} proves the corollary.
\end{proof}

\subsection*{Disclosure of Generative AI Use}

The proofs of Theorems~\ref{thm:chordal} and~\ref{thm:strongchordal} were developed from initial reductions and proof drafts generated through interaction with ChatGPT (OpenAI; GPT-5.6 Sol). ChatGPT (GPT-6 Astra) and Claude (Anthropic; Claude Opus 5.5) were subsequently used to refine the exposition.
ChatGPT also assisted with LaTeX/TikZ preparation. The author takes full
responsibility for the final manuscript.

\bibliographystyle{plainurl}
\begingroup
\small
\bibliography{eds_chordal}

@inproceedings{MH77,
  author       = {S. L. Mitchell and
                  S. T. Hedetniemi},
  title        = {Edge domination in trees},
  booktitle    = {Proceedings of the Eighth Southeastern Conference on Combinatorics,
                  Graph Theory and Computing},
  series       = {Congressus Numerantium},
  volume       = {19},
  pages        = {489--509},
  publisher    = {Utilitas Mathematica},
  year         = {1977},
  bibsource    = {https://www.researchgate.net/profile/Stephen-Hedetniemi/publication/246012399_Edge_domination_in_trees/links/5510bcf90cf20352196c70c9/Edge-domination-in-trees.pdf}
}

@article{RP88,
  author       = {M. B. Richey and
                  R. Gary Parker},
  title        = {Minimum-maximal matching in series-parallel graphs},
  journal      = {Eur. J. Oper. Res.},
  volume       = {33},
  number       = {1},
  pages        = {98--105},
  year         = {1988},
  url          = {https://doi.org/10.1016/0377-2217(88)90258-5},
  doi          = {10.1016/0377-2217(88)90258-5},
  bibsource    = {https://www.sciencedirect.com/science/article/abs/pii/0377221788902585}
}

@article{KR03,
  author       = {Daniel Kobler and
                  Udi Rotics},
  title        = {Edge dominating set and colorings on graphs with fixed clique-width},
  journal      = {Discret. Appl. Math.},
  volume       = {126},
  number       = {2-3},
  pages        = {197--221},
  year         = {2003},
  url          = {https://doi.org/10.1016/S0166-218X(02)00198-1},
  doi          = {10.1016/S0166-218X(02)00198-1},
  bibsource    = {https://www.sciencedirect.com/science/article/pii/S0166218X02001981}
}

@inproceedings{DE08,
  author       = {Marc Demange and
                  Tinaz Ekim},
  editor       = {Manindra Agrawal and
                  Ding{-}Zhu Du and
                  Zhenhua Duan and
                  Angsheng Li},
  title        = {Minimum Maximal Matching Is {NP}-Hard in Regular Bipartite Graphs},
  booktitle    = {Theory and Applications of Models of Computation, 5th International
                  Conference, {TAMC} 2008, Xi'an, China, April 25-29, 2008. Proceedings},
  series       = {Lecture Notes in Computer Science},
  volume       = {4978},
  pages        = {364--374},
  publisher    = {Springer},
  year         = {2008},
  url          = {https://doi.org/10.1007/978-3-540-79228-4_32},
  doi          = {10.1007/978-3-540-79228-4_32},
  bibsource    = {https://link.springer.com/chapter/10.1007/978-3-540-79228-4_32}
}

@article{YG80,
  author       = {M. Yannakakis and
                  F. Gavril},
  title        = {Edge Dominating Sets in Graphs},
  journal      = {{SIAM} J. Appl. Math.},
  volume       = {38},
  number       = {3},
  pages        = {364--372},
  year         = {1980},
  url          = {https://doi.org/10.1137/0138030},
  doi          = {10.1137/0138030},
  bibsource    = {https://epubs.siam.org/doi/10.1137/0138030}
}

@book{GJ79,
  author       = {M. R. Garey and
                  David S. Johnson},
  title        = {Computers and Intractability: A Guide to the Theory of
                  {NP}-Completeness},
  publisher    = {W. H. Freeman},
  year         = {1979},
  isbn         = {0-7167-1044-7},
  bibsource    = {dblp computer science bibliography, https://dblp.org}
}

@article{GJ78,
  author       = {M. R. Garey and
                  David S. Johnson},
  title        = {{``Strong''} {NP}-Completeness Results: Motivation, Examples, and
                  Implications},
  journal      = {J. {ACM}},
  volume       = {25},
  number       = {3},
  pages        = {499--508},
  year         = {1978},
  url          = {https://doi.org/10.1145/322077.322090},
  doi          = {10.1145/322077.322090},
  bibsource    = {dblp computer science bibliography, https://dblp.org}
}

@article{Farber83,
  author       = {Martin Farber},
  title        = {Characterizations of strongly chordal graphs},
  journal      = {Discret. Math.},
  volume       = {43},
  number       = {2-3},
  pages        = {173--189},
  year         = {1983},
  url          = {https://doi.org/10.1016/0012-365X(83)90154-1},
  doi          = {10.1016/0012-365X(83)90154-1},
  bibsource    = {dblp computer science bibliography, https://dblp.org}
}

@article{Farber84,
  author       = {Martin Farber},
  title        = {Domination, independent domination, and duality in strongly chordal
                  graphs},
  journal      = {Discret. Appl. Math.},
  volume       = {7},
  number       = {2},
  pages        = {115--130},
  year         = {1984},
  url          = {https://doi.org/10.1016/0166-218X(84)90061-1},
  doi          = {10.1016/0166-218X(84)90061-1},
  bibsource    = {dblp computer science bibliography, https://dblp.org}
}

@article{CFT93,
  author       = {Gerard J. Chang and
                  Martin Farber and
                  Zsolt Tuza},
  title        = {Algorithmic Aspects of Neighborhood Numbers},
  journal      = {{SIAM} J. Discret. Math.},
  volume       = {6},
  number       = {1},
  pages        = {24--29},
  year         = {1993},
  url          = {https://doi.org/10.1137/0406002},
  doi          = {10.1137/0406002}
}

@article{BCD98,
  author       = {Andreas Brandst{\"{a}}dt and
                  Victor Chepoi and
                  Feodor F. Dragan},
  title        = {The Algorithmic Use of Hypertree Structure and Maximum Neighbourhood
                  Orderings},
  journal      = {Discret. Appl. Math.},
  volume       = {82},
  number       = {1-3},
  pages        = {43--77},
  year         = {1998},
  url          = {https://doi.org/10.1016/S0166-218X(97)00125-X},
  doi          = {10.1016/S0166-218X(97)00125-X},
  bibsource    = {https://pageperso.lis-lab.fr/~victor.chepoi/AlgUseMN.pdf}
}

@article{HK93,
  author       = {Joseph Douglas Horton and
                  Kyriakos Kilakos},
  title        = {Minimum Edge Dominating Sets},
  journal      = {{SIAM} J. Discret. Math.},
  volume       = {6},
  number       = {3},
  pages        = {375--387},
  year         = {1993},
  url          = {https://doi.org/10.1137/0406030},
  doi          = {10.1137/0406030},
  bibsource    = {dblp computer science bibliography, https://dblp.org}
}

@article{S95,
  author       = {Anand Srinivasan and
                  K. Madhukar and
                  P. Nagavamsi and
                  C. Pandu Rangan and
                  Maw{-}Shang Chang},
  title        = {Edge Domination on Bipartite Permutation Graphs and Cotriangulated
                  Graphs},
  journal      = {Inf. Process. Lett.},
  volume       = {56},
  number       = {3},
  pages        = {165--171},
  year         = {1995},
  url          = {https://doi.org/10.1016/0020-0190(95)94093-8},
  doi          = {10.1016/0020-0190(95)94093-8},
  bibsource    = {dblp computer science bibliography, https://dblp.org}
}

@article{HC95,
  author       = {Shiow{-}Fen Hwang and
                  Gerard J. Chang},
  title        = {The edge domination problem},
  journal      = {Discuss. Math. Graph Theory},
  volume       = {15},
  number       = {1},
  pages        = {51--57},
  year         = {1995},
  url          = {https://doi.org/10.7151/dmgt.1006},
  doi          = {10.7151/dmgt.1006},
  bibsource    = {DBLP; author order checked against the original article}
}

@article{Manlove,
  author       = {David F. Manlove},
  title        = {On the Algorithmic Complexity of Twelve Covering and Independence
                  Parameters of Graphs},
  journal      = {Discret. Appl. Math.},
  volume       = {91},
  number       = {1-3},
  pages        = {155--175},
  year         = {1999},
  url          = {https://doi.org/10.1016/S0166-218X(98)00147-4},
  doi          = {10.1016/S0166-218X(98)00147-4},
  bibsource    = {dblp computer science bibliography, https://dblp.org}
}

@article{Bertossi,
  author       = {Alan A. Bertossi},
  title        = {Dominating Sets for Split and Bipartite Graphs},
  journal      = {Inf. Process. Lett.},
  volume       = {19},
  number       = {1},
  pages        = {37--40},
  year         = {1984},
  url          = {https://doi.org/10.1016/0020-0190(84)90126-1},
  doi          = {10.1016/0020-0190(84)90126-1},
  bibsource    = {dblp computer science bibliography, https://dblp.org}
}

@article{LKT,
  author       = {Chin Lung Lu and
                  Ming{-}Tat Ko and
                  Chuan Yi Tang},
  title        = {Perfect edge domination and efficient edge domination in graphs},
  journal      = {Discret. Appl. Math.},
  volume       = {119},
  number       = {3},
  pages        = {227--250},
  year         = {2002},
  url          = {https://doi.org/10.1016/S0166-218X(01)00198-6},
  doi          = {10.1016/S0166-218X(01)00198-6},
  bibsource    = {dblp computer science bibliography, https://dblp.org}
}

@article{Dean,
  author       = {Brian C. Dean and
                  Sandra Mitchell Hedetniemi and
                  Stephen T. Hedetniemi and
                  Jason Lewis and
                  Alice A. McRae},
  title        = {Matchability and k-maximal matchings},
  journal      = {Discret. Appl. Math.},
  volume       = {159},
  number       = {1},
  pages        = {15--22},
  year         = {2011},
  url          = {https://doi.org/10.1016/j.dam.2010.09.006},
  doi          = {10.1016/j.dam.2010.09.006},
  bibsource    = {dblp computer science bibliography, https://dblp.org}
}

@article{FG,
  author       = {D. R. Fulkerson and
                  O. A. Gross},
  title        = {Incidence matrices and interval graphs},
  journal      = {Pac. J. Math.},
  volume       = {15},
  number       = {3},
  pages        = {835--855},
  year         = {1965},
  url          = {https://doi.org/10.2140/pjm.1965.15.835},
  doi          = {10.2140/pjm.1965.15.835},
  bibsource    = {https://msp.org/pjm/1965/15-3/pjm-v15-n3-p11-s.pdf}
}

@article{Milanic22,
  author       = {Martin Milani{\v{c}}},
  title        = {An open problem on edge domination in graphs},
  journal      = {Indian J. Discret. Math.},
  volume       = {8},
  number       = {1},
  pages        = {43--46},
  year         = {2022},
  url          = {http://ijdm.adma.co.in}
}

@inproceedings{EGW01,
  author       = {Wolfgang Espelage and
                  Frank Gurski and
                  Egon Wanke},
  editor       = {Andreas Brandst{\"{a}}dt and
                  Van Bang Le},
  title        = {How to Solve {NP}-hard Graph Problems on Clique-Width Bounded Graphs
                  in Polynomial Time},
  booktitle    = {Graph-Theoretic Concepts in Computer Science, 27th International
                  Workshop, {WG} 2001, Boltenhagen, Germany, June 14-16, 2001,
                  Proceedings},
  series       = {Lecture Notes in Computer Science},
  volume       = {2204},
  pages        = {117--128},
  publisher    = {Springer},
  year         = {2001},
  url          = {https://doi.org/10.1007/3-540-45477-2_12},
  doi          = {10.1007/3-540-45477-2_12}
}

@article{SawadaS03,
  author       = {Joe Sawada and
                  Jeremy P. Spinrad},
  title        = {From a simple elimination ordering to a strong elimination ordering in
                  linear time},
  journal      = {Inf. Process. Lett.},
  volume       = {86},
  number       = {6},
  pages        = {299--302},
  year         = {2003},
  url          = {https://doi.org/10.1016/S0020-0190(03)00228-X},
  doi          = {10.1016/S0020-0190(03)00228-X},
  bibsource    = {dblp computer science bibliography, https://dblp.org}
}

@article{CourcelleO00,
  author       = {Bruno Courcelle and
                  Stephan Olariu},
  title        = {Upper bounds to the clique width of graphs},
  journal      = {Discret. Appl. Math.},
  volume       = {101},
  number       = {1-3},
  pages        = {77--114},
  year         = {2000},
  url          = {https://doi.org/10.1016/S0166-218X(99)00184-5},
  doi          = {10.1016/S0166-218X(99)00184-5}
}

@article{CorneilLS81,
  author       = {Derek G. Corneil and
                  H. Lerchs and
                  L. Stewart Burlingham},
  title        = {Complement reducible graphs},
  journal      = {Discret. Appl. Math.},
  volume       = {3},
  number       = {3},
  pages        = {163--174},
  year         = {1981},
  url          = {https://doi.org/10.1016/0166-218X(81)90013-5},
  doi          = {10.1016/0166-218X(81)90013-5}
}
\endgroup
\end{document}